\documentclass[hyper]{JHEP} 

\usepackage{epsfig}

\newcommand\fverb{\setbox\pippobox=\hbox\bgroup\verb}

\newcommand\fverbdo{\egroup\medskip\noindent%

            \fbox{\unhbox\pippobox}\ }

\newcommand\fverbit{\egroup\item[\fbox{\unhbox\pippobox}]}

\newbox\pippobox

\title{Unstable D-brane in Torsional Newton-Cartan Background}
\author{J. Kluso\v{n}\\
Department of
Theoretical Physics and Astrophysics\\
Faculty of Science, Masaryk University\\
Kotl\'{a}\v{r}sk\'{a} 2, 611 37, Brno\\
Czech Republic\\
E-mail: \email{klu@physics.muni.cz}} \preprint{}

 \abstract{This paper is devoted to the construction of
 unstable D-brane action in torsional Newton-Cartan background through T-duality
along null direction. We determine corresponding equations of motion
and analyze their solution that corresponds to lower dimensional non-relativistic
D(p-1)-brane. We also find Hamiltonian for unstable Dp-brane and study tachyon
vacuum solutions that can be interpreted as gas of non-relativistic strings.}

\def\hv{\hat{v}}

\def\hl{\hat{l}}

\def\hgamma{\hat{\gamma}}

\def\balpha{\bar{\alpha}}
\def\bbeta{\bar{\beta}}

\def\bA{\mathbf{A}}

\def\ttau{\tilde{\tau}}

\def\tT{\tilde{T}}

\def\be{\begin{equation}}

\def\halpha{\hat{\alpha}}
\def\hbeta{\hat{\beta}}
\def\ee{\end{equation}}

\def\bea{\begin{eqnarray}}
\def\bh{\bar{h}}
\def\hi{\hat{i}}
\def\hj{\hat{j}}
\def\eea{\end{eqnarray}}

\def\hdelta{\hat{\delta}}

\def\mH{\mathcal{H}}

\def\hbA{\hat{\bA}}

\newcommand{\mK}{\mathcal{K}}
\def\hbA{\hat{\bA}}

\newcommand{\mG}{\mathcal{G}}

\newcommand{\hk}{\hat{k}}

\def \bA{\mathbf{A}}

\def\bgamma{\bar{\gamma}}
\def\bdelta{\bar{\delta}}

\newcommand{\ba}{\mathbf{a}}

\def\pb #1{\left\{#1\right\}}

\begin{document}
\section{Introduction and Summary }
Newton Cartan (NC) gravity \cite{Cartan:1923zea}, which is covariant formulation of non-relativistic gravity, has been intensively studied in the past few years from different points of view. For example, torsionful generalization of NC geometry
has an important place in the study of non-relativistic aspects of string theory and holography. Torsionful generalization NC  which has non-exact clock form, was firstly observed as the boundary
geometry in the context of Lifshitz holography \cite{Christensen:2013lma,Christensen:2013rfa,Hartong:2014oma}. There is also interesting relation between NC gravity and Ho\v{r}ava-Lifshitz gravity \cite{Horava:2009uw} that was found in \cite{Hartong:2015zia,Hartong:2016yrf}. Finally, there is also great interest in the analysis of non-relativistic string theory
\cite{Gomis:2000bd,Danielsson:2000gi}  and its covariant version that was studied in
\cite{Andringa:2012uz,Harmark:2017rpg,Bergshoeff:2018yvt,Kluson:2018egd,Kluson:2018grx,
	Harmark:2018cdl,Gomis:2019zyu,Gallegos:2019icg,Bergshoeff:2019pij,Harmark:2019upf}.

At present there are two versions of non-relativistic string theories in NC background. The first one
corresponds to strings on torsional NC background
\cite{Harmark:2017rpg,Harmark:2018cdl,Harmark:2019upf} which is basically defined
as null reduction of relativistic string. The second one is known as
stringy NC gravity \cite{Andringa:2012uz,Bergshoeff:2018yvt,Bergshoeff:2019pij} and it
is defined as the special limit from relativistic string theory
\cite{Andringa:2012uz}. This limit can be considered as a generalization of the analysis presented \cite{Bergshoeff:2015uaa} when the point particle probe is replaced by fundamental string.
The beta function of the non-relativistic string in stringy NC background was determined in \cite{Gomis:2019zyu} that leads to the dynamical equations of corresponding
stringy NC gravity. In case of torsional NC gravity the same result was derived
in \cite{Gallegos:2019icg}.
Then it was shown recently in \cite{Harmark:2019upf} that these two non-relativistic string theories can be mapped each other.

These results show that non-relativistic string theories are very important and certainly deserve further study. For example, in our previous paper
\cite{Kluson:2019avy} we introduced non-relativistic D-branes in torsional NC background through dimensional reduction along null direction that can be interpreted as T-duality
transformation of relativistic D(p+1)-brane along null direction
\footnote{For extended review of D-branes and their T-duality transformations, see for example  \cite{Simon:2011rw}.}. Performing this T-duality transformation we derived an action for non-relativistic Dp-brane in torsional NC background and we studied its properties. Explicitly, we derived its Hamiltonian form and we also analyzed how its transforms under T-duality along spatial direction.

In this paper we would like to extend this analysis to the case of unstable D-branes which are objects known from relativistic string theories and where they have very important place
\footnote{For review and extensive list of references, see
\cite{Sen:2004nf}.}. Relativistic unstable D-brane is characterized by presence of the tachyon field $T$ with the potential $V(T)$ that is even function of $T$ and that has global minimum for $T_{min}=\pm \infty$ where $V(T_{min})=0$ where unstable D-brane disappears while local unstable maximum at $T_{max}=0, V(T_{max})=1$. Further, this D-brane breaks target space supersymmetry completely. These objects are described by non-BPS D-brane effective actions \cite{Sen:1999md,Kluson:2000iy,Bergshoeff:2000dq}.
This action will be the starting point for the definition of unstable D-brane in torsional NC background when we closely follow analysis performed in
\cite{Kluson:2018egd}. Explicitly, we consider unstable D(p+1)-brane in the   background with the null isometry in the form that was suggested in \cite{Harmark:2017rpg} and perform double dimensional reduction along null direction.  We obtain an action for unstable Dp-brane in torsional NC background and we study its property. Explicitly, we focus on equations of motion and their solution in the form of the tachyon kink. It was shown in
very nice paper \cite{Sen:2003tm} in the context of non-BPS Dirac-Born-Infeld (DBI) action that  fluctuations around  tachyon kink solution obey the equations of motion that can be derived an action for codimension one stable D-brane. Following the same analysis in case of the non-relativistic unstable Dp-brane we obtain that the tachyon condensation again leads to the lower dimensional non-relativistic D(p-1)-brane in torsional NC background. This fact can be considered
as nice consistency check of the theory.

As the next step we would like to answer the question of the behaviour of non-relativistic unstable Dp-brane at the tachyon vacuum $T_{min}=\infty$. Since the tachyon effective action is multiplied by $V$ we see that at the tachyon vacuum this action vanishes and hence it is difficult to analyze fluctuations around its ground state. Then it was shown in
\cite{Sen:2003bc,Sen:2000kd} that more natural is to switch to Hamiltonian formalism and study canonical equations of motion at the point $T_{min}=\infty$. It was shown there that the canonical equations of motion at the $T=T_{min}$ describe gas of fundamental strings. We study the same problem in case of non-relativistic D-brane  in torsional NC background. We firstly determine Hamiltonian for unstable D-brane through the null dimensional reduction of the Hamiltonian of relativistic unstable D(p+1)-brane. Then we determine canonical equations of motion and study  their properties at the tachyon vacuum. We show that the tachyon vacuum solution describes fundamental non-relativistic string in torsionful NC background.

Let us outline our results. We determined form of unstable non-relativistic Dp-brane in
torsionful NC background through null dimensional reduction of relativistic unstable D(p+1)-brane. We derived equations of motions and we studied tachyon kink solution. We argued that it correspond to non-relativistic Dp-brane. We also determined Hamiltonian for this unstable Dp-brane and studied its behaviour at tachyon vacuum. We showed that corresponding equations of motion describe gas of fundamental non-relativistic strings.
These results again confirm that non-relativistic unstable D-brane shares the same properties with the relativistic one can be considered as a consistency check of the non-relativistic unstable D-brane action.

This paper is organized as follows. In the next section (\ref{second}) we obtain an action for non-relativistic unstable D-brane in torsional NC background. In section (\ref{third}) we derive equations of motion and analyze their solutions in the form of the tachyon kink. In section (\ref{fourth}) we focus on Hamiltonian for unstable D-brane and study
canonical equations of motion at the tachyon vacuum.

\section{Non-Relativistic Unstable Dp-brane}\label{second}
In this section we derive unstable Dp-brane in torsional NC background through double
dimensional reduction along null direction. We start with the relativistic
unstable  D(p+1)-brane whose action has the form
 \cite{Sen:1999md,Kluson:2000iy,Bergshoeff:2000dq}
\begin{eqnarray}
& &S_{NBPS}=-\tT_{p+1}\int d^{p+2}\xi e^{-\phi'} V(T)\sqrt{-\det \tilde{\bA}_{\alpha\beta}} \ ,
\nonumber \\
& &\tilde{\bA}_{\alpha\beta}=G_{MN}\partial_\alpha x^M\partial_\beta x^N+B_{MN}
\partial_\alpha x^M\partial_\beta x^N+l_s^2 F_{\alpha\beta}+l_s^2 \partial_\alpha T
\partial_\beta T \ , \nonumber \\
\end{eqnarray}
where $T$ is tachyon, $V(T)$ is tachyon potential that is even function with the property that
$V(T_{min})=0$ for $T_{min}=\pm \infty$ and $V(T_{max})=1$. Further, $\phi' ,G_{MN},B_{MN},M,N=0,1,\dots,9 $ are background dilaton, metric and NSNS two form field, respectively. We parameterize the world volume of D(p+1)-brane with coordinates $\xi^\alpha,\alpha=0,1,\dots,p+1$. The remaining world-volume fields are $x^M(\xi)$ that parameterize an embedding of D(p+1)-brane in target spacetime and gauge field $A_\alpha$ with the field strength $F_{\alpha\beta}=\partial_\alpha A_\beta-\partial_\beta A_\alpha$. Finally, $l_s$ is string length and $\tT_{p+1}$ is unstable D(p+1)-brane tension  $\tT_{p+1}=\frac{\sqrt{2}}{l_s^{p+2}}=
\sqrt{2}T^{BPS}_{p+1}$, where $T_{BPS}$ is tension of the stable D(p+1)-brane.

Let us consider this D(p+1)-brane in the background with null isometry where the line element has the form \cite{Harmark:2017rpg,Harmark:2018cdl,Harmark:2019upf}
\begin{equation}\label{Nullback}
ds^2=g_{MN}dx^M dx^N=2\tau (du-m)+h_{\mu\nu}dx^\mu dx^\nu \ ,
\tau=\tau_\mu dx^\mu \ , m=m_\mu dx^\mu \ ,
\end{equation}
where $\det h_{\mu\nu}=0$ and where the background possesses an isometry $
u\rightarrow u+\epsilon,\epsilon=\mathrm{const}$.
Now we presume that D(p+1)-brane is extended along $u-$direction so that we can partially fix the gauge
\begin{equation}
u=\xi^{p+1}
\end{equation}
and we further presume that all world-volume fields do not depend on $\xi^{p+1}$. Let $\halpha,\hbeta$ denote
remaining world-volume  coordinates $\halpha=0,1,\dots,p$. Then the matrix $\tilde{\bA}_{\alpha\beta}$
has the form
\begin{eqnarray}
\tilde{\bA}_{\alpha\beta}=
\left(\begin{array}{cc}
\hbA_{\halpha\hbeta} &
\tau_{\halpha}+l_s^2 \partial_{\halpha}A_{(p+1)} \\
\tau_{\hbeta}-l_s^2\partial_{\hbeta}A_{(p+1)} & 0 \\
\end{array}\right) \ ,
\nonumber \\
\end{eqnarray}
where $\hbA_{\halpha\hbeta}$ has the form
\begin{equation}
\hbA_{\halpha\hbeta}=\bh_{\halpha\hbeta}+l_s^2 F_{\halpha\hbeta}+
l_s^2 \partial_{\halpha}T \partial_{\hbeta}T \
\end{equation}
and where $\bh_{\halpha\beta}=\bh_{\mu\nu}\partial_{\halpha}x^\mu\partial_{\hbeta}x^\nu \ ,
\quad \bh_{\mu\nu}=h_{\mu\nu}-m_\mu\tau_\nu-m_\nu\tau_\mu$.

Let   $\hbA^{\halpha\hbeta}$ is an inverse matrix to $\hbA_{\halpha\beta}$  so that
\begin{equation}
\hbA_{\halpha\hbeta}\hbA^{\hbeta\hgamma}=
\delta_{\halpha}^{\hgamma} \ .
\end{equation}
Then it is easy to find an   action for non-relativistic unstable Dp-brane in torsional NC background in the form
\begin{equation}\label{actnon}
S=-\tT_p\int d^{p+1}\xi V(T)e^{-\phi}
\sqrt{\det\hbA}
\sqrt{(\tau_{\halpha}-\partial_{\halpha}\eta)\hbA^{\halpha\hbeta}
(\tau_{\hbeta}+\partial_{\hbeta}\eta)} \ ,
\end{equation}
where we finally  defined $\tT_p$ and $e^{-\phi}$ through the relation
\begin{equation}
\tT_p e^{-\phi}=\int du \tT_{p+1}e^{-\phi'} \ .
\end{equation}
Since the action described unstable Dp-brane the natural question is to analyze
solutions of the equations of motion that follow from (\ref{actnon}) and
that correspond to some stable configurations. Such a solution is known as tachyon
kink and we will study its properties in the next section.
\section{Tachyon Kink Solution}\label{third}
We would like to study equations of motion for non-relativistic unstable Dp-brane
in torsional NC background that can be derived from (\ref{actnon}).  The variation of the action (\ref{actnon}) with respect to $\eta$ gives equation of motion for $\eta$
\begin{eqnarray}\label{eqeta}
\partial_{\halpha}\left[\frac{V(T)e^{-\phi}\sqrt{\det\hbA}\hbA^{\halpha
		\hbeta}\partial_{\hbeta}\eta}
	{\sqrt{(\tau_{\halpha}-\partial_{\halpha}\eta)\hbA^{\halpha\hbeta}(\tau_{\hbeta}+
			\partial_{\hbeta}\eta)}}\right]=0 \ . \nonumber \\
		\end{eqnarray}
Further, variation of (\ref{actnon}) with respect to $T$ gives 	
\begin{eqnarray}\label{eqT}
& &-e^{-\phi}\frac{dV}{dT}\sqrt{\det\hbA}\sqrt{(\tau_{\halpha}-\partial_{\halpha}\eta)\hbA^{\halpha\hbeta}
	(\tau_{\hbeta}+\partial_{\hbeta}\eta)}+\nonumber \\
& &+l_s^2
\partial_{\halpha}\left[V(T)e^{-\phi}\partial_{\hbeta}T
\hbA^{\hbeta\halpha}_S\sqrt{\det\hbA}\sqrt{(\tau_{\halpha}-\partial_{\halpha}\eta)\hbA^{\halpha\hbeta}
	(\tau_{\hbeta}+\partial_{\hbeta}\eta)}\right]-
\nonumber \\
& &-\frac{l_s^2}{2}\partial_{\hgamma}\left[
\frac{e^{-\phi}V(T)\sqrt{\det\hbA }}
{\sqrt{(\tau_{\halpha}-\partial_{\halpha}\eta)\hbA^{\halpha\hbeta}
		(\tau_{\hbeta}+\partial_{\hbeta}\eta)}}
(\tau_{\halpha}-\partial_{\halpha}\eta)\hbA^{\halpha\hgamma}\partial_{\hdelta}
T\hbA^{\hdelta\hbeta}(\tau_{\hbeta}+\partial_{\hbeta}\eta)\right]-
\nonumber \\
& &-\frac{l_s^2}{2}\partial_{\hdelta}\left[
\frac{e^{-\phi}V(T)\sqrt{\det\hbA}}
{\sqrt{(\tau_{\halpha}-\partial_{\halpha}\eta)\hbA^{\halpha\hbeta}
		(\tau_{\hbeta}+\partial_{\hbeta}\eta)}}
(\tau_{\halpha}-\partial_{\halpha}\eta)\hbA^{\halpha\hgamma}\partial_{\hgamma}
T\hbA^{\hdelta\hbeta}(\tau_{\hbeta}+\partial_{\hbeta}\eta)\right]=0 \ .
\nonumber \\
\end{eqnarray}
Further, equation of motion with respect to $A_{\halpha}$ take the form
\begin{eqnarray}\label{eqA}
& &\partial_{\hbeta}[e^{-\phi}V(T)\bA^{\halpha\hbeta}_A\sqrt{\det\hbA}
\sqrt{(\tau_{\halpha}-\partial_{\halpha}\eta)\hbA^{\halpha\hbeta}
	(\tau_{\hbeta}+\partial_{\hbeta}\eta)}]-
\nonumber \\
& &-\frac{1}{2}\partial_{\hbeta}\left[Ve^{-\phi}\frac{\sqrt{\det\hbA}}
{\sqrt{(\tau_{\halpha}-\partial_{\halpha}\eta)\hbA^{\halpha\hbeta}
		(\tau_{\hbeta}+\partial_{\hbeta}\eta)}}	
(\tau_{\hgamma}-\partial_{\hgamma}\eta)\hbA^{\hgamma\hbeta}\hbA^{\halpha\hdelta}
(\tau_{\hdelta}+\partial_{\hdelta}\eta)\right]+\nonumber \\
& &+\frac{1}{2}\partial_{\hbeta}\left[e^{-\phi}\frac{\sqrt{\det\hbA}}{
\sqrt{(\tau_{\halpha}-\partial_{\halpha}\eta)\hbA^{\halpha\hbeta}
	(\tau_{\hbeta}+\partial_{\hbeta}\eta)}}
(\tau_{\hgamma}-\partial_{\hgamma}\eta)\hbA^{\hgamma\halpha}
\hbA^{\hbeta\hdelta}(\tau_{\hdelta}+\partial_{\hdelta}\eta)\right]=0 \ .
\nonumber \\
\end{eqnarray}
where we defined symmetric and anti-symmetric combination of matrix as
\begin{equation}
\bA^{\balpha\bbeta}_S=\frac{1}{2}(\bA^{\balpha\bbeta}+\bA^{\bbeta\balpha}) \ , \quad
\bA^{\balpha\bbeta}_A=\frac{1}{2}(\bA^{\balpha\bbeta}-\bA^{\bbeta\balpha}) \ .
\end{equation}
Finally we determine equation of motion for $x^\mu$ in the form
\begin{eqnarray}\label{eqxmu}
& & V(T)\partial_\mu\phi e^{-\phi}
\sqrt{\det\hbA}
\sqrt{(\tau_{\halpha}-\partial_{\halpha}\eta)\hbA^{\halpha\hbeta}
	(\tau_{\hbeta}+\partial_{\hbeta}\eta)}-\nonumber \\
& &-\frac{1}{2} V(T)e^{-\phi}
\partial_\mu \bh_{\rho\sigma}\partial_{\halpha}x^\rho
\partial_{\hbeta}x^\sigma \hbA^{\hbeta\halpha}\sqrt{\det\hbA}
\sqrt{(\tau_{\halpha}-\partial_{\halpha}\eta)\hbA^{\halpha\hbeta}
	(\tau_{\hbeta}+\partial_{\hbeta}\eta)}
+\nonumber \\
& &+\partial_{\halpha}\left[ e^{-\phi}V
\bh_{\mu\nu}\partial_{\hbeta}x^\nu \hbA^{\hbeta\halpha}_S\sqrt{\det\hbA}
\sqrt{(\tau_{\halpha}-\partial_{\halpha}\eta)\hbA^{\halpha\hbeta}
	(\tau_{\hbeta}+\partial_{\hbeta}\eta)}\right]+\nonumber \\
& &+\partial_{\halpha}\left[ V(T)e^{-\phi}\sqrt{\det\hbA}\tau_\mu
\frac{\hbA^{\halpha\hbeta}_S\tau_{\hbeta}+\hbA^{\halpha\hbeta}_A\partial_{\hbeta}\eta}{
\sqrt{(\tau_{\halpha}-\partial_{\halpha}\eta)\hbA^{\halpha\hbeta}
	(\tau_{\hbeta}+\partial_{\hbeta}\eta)}}\right]+ \nonumber \\
& &+\frac{1}{2} V(T)e^{-\phi}\frac{\sqrt{\det\hbA}}
{\sqrt{(\tau_{\halpha}-\partial_{\halpha}\eta)\hbA^{\halpha\hbeta}
		(\tau_{\hbeta}+\partial_{\hbeta}\eta)}}
(\tau_{\hgamma}-\partial_{\hgamma}\eta)\hbA^{\hgamma\halpha}
\partial_\mu \bh_{\rho\sigma}\partial_{\halpha}x^\rho
\partial_{\hbeta}x^\sigma \hbA^{\hbeta\hdelta}(\tau_{\hdelta}+
\partial_{\hdelta}\eta)-\nonumber \\ \nonumber \\
& &-\frac{1}{2}\partial_{\hgamma}\left[
V(T)e^{-\phi}\frac{\sqrt{\det\hbA}}{\sqrt{(\tau_{\halpha}-\partial_{\halpha}\eta)\hbA^{\halpha\hbeta}
		(\tau_{\hbeta}+\partial_{\hbeta}\eta)}}(\tau_{\halpha}-\partial_{\halpha}\eta)
\hbA^{\halpha\hgamma}\bh_{\mu\nu}\partial_{\hdelta} x^\nu \hbA^{\hdelta
	\hbeta}(\tau_{\hbeta}+\partial_{\hbeta}\eta)\right]-\nonumber \\
& &-\frac{1}{2}\partial_{\hdelta}\left[
V(T)e^{-\phi}\frac{\sqrt{\det\hbA}}{\sqrt{(\tau_{\halpha}-\partial_{\halpha}\eta)\hbA^{\halpha\hbeta}
		(\tau_{\hbeta}+\partial_{\hbeta}\eta)}}(\tau_{\halpha}-\partial_{\halpha}\eta)
\hbA^{\halpha\hgamma}\partial_{\hgamma}x^\nu\bh_{\nu\mu} \hbA^{\hdelta
	\hbeta}(\tau_{\hbeta}+\partial_{\hbeta}\eta)\right]=0 \ . \nonumber \\
\end{eqnarray}
Let us now study tachyon condensation on this D-brane following analysis presented in
\cite{Sen:2003tm}.
We select one world-volume coordinate $\xi^{p}=y$ and label remaining ones with
$\xi^{\balpha},\balpha,\bbeta=0,\dots,p-1$. Let us now presume tachyon kink solution in the form
\begin{equation}
T(y,\xi^{\balpha})=f(a(y-t(\xi^{\balpha}))) \ ,
\end{equation}
where $f(u)$ satisfies following conditions
\begin{equation}
f(-u)=-f(u) \ , \quad   f'(u)>0 \ , \forall u \ , \quad f(\pm\infty)=\pm \infty \ .
\end{equation}
We further presume that all remaining fields depend on $\xi^{\balpha}$ only so that
\begin{equation}\label{ans}
x^\mu(y,\xi^{\balpha})=x^\mu(\xi^{\balpha}) \ ,\quad
A_y=0 \ , \quad  A_{\balpha}(y,\xi^{\balpha})=a_{\balpha}(\xi^{\balpha}) \ , \quad
\eta(y,\xi^{\balpha})=\eta(\xi^{\balpha}) \
\end{equation}
so that the matrix $\hbA_{\halpha\hbeta}$ has the form
\begin{equation}
\hbA_{\balpha\bbeta}=\ba_{\balpha\bbeta}+a^2 f'^2\partial_{\balpha}t\partial_{\bbeta}t \ , \quad \hbA_{\balpha y}=-a^2 f'^2\partial_{\balpha}t \ , \quad
\hbA_{y\bbeta}=-a^2 f'^2\partial_{\bbeta}t \ , \quad  \hbA_{yy}=a^2 f'^2 \ ,
\end{equation}
where
\begin{equation}
\ba_{\balpha\bbeta}=\bh_{\mu\nu}\partial_{\balpha}x^{\mu}\partial_{\bbeta}x^\nu+l_s^2
F_{\balpha\bbeta} \  .
\end{equation}
Then it is easy to see that the  inverse matrix $\hbA^{\halpha\hbeta}$  has the form
\begin{eqnarray}
& &\hbA^{yy}=\frac{1}{a^2 f'^2}+\partial_{\bgamma}t \ba^{\bgamma\bdelta}\partial_{\bdelta}t \ ,
\nonumber \\
& & \hbA^{y\bbeta}=\partial_{\bdelta}t \ba^{\bdelta \bbeta} \ ,
\quad
\hbA^{\balpha y}=\ba^{\balpha\bgamma}\partial_{\bgamma}t \ , \quad   \hbA^{\balpha\bbeta}=
\ba^{\balpha\bbeta} \ , \nonumber \\
\end{eqnarray}
where $\ba^{\balpha\bbeta}$ is matrix inverse to $\ba_{\balpha\bbeta}$.
To proceed further we observe that for the ansatz given in (\ref{ans}) we have
\begin{eqnarray}
& &\det \hbA=a^2 f'^2 \det\ba
 \ ,  \nonumber \\
& &(\tau_{\halpha}-\partial_{\halpha}\eta)\hbA^{\halpha\hbeta}
(\tau_{\hbeta}+\partial_{\hbeta}\eta)=(\tau_{\balpha}-\partial_{\balpha}\eta)
\ba^{\balpha\bbeta}(\tau_{\bbeta}+\partial_{\bbeta}\eta)
\end{eqnarray}
and also
\begin{equation}
\partial_{\halpha}T \hbA^{\halpha\bbeta}_S=0 \ ,
\quad \partial_{\halpha}T\hbA^{\halpha y}_S=\frac{1}{af'} \ .
\end{equation}
Inserting these results into (\ref{eqeta}) we obtain that it has the form
\begin{equation}\label{eqetapre}
V(T)af'\partial_{\balpha}\left[\frac{e^{-\phi}\sqrt{\det\ba}\ba^{\balpha\bbeta}_S\partial_{\bbeta}\eta}{\sqrt{(\tau_{\balpha}-\partial_{\balpha}\eta)\ba^{\balpha\bbeta}
(\tau_{\bbeta}+\partial_{\bbeta}\eta)}}\right]=0
\end{equation}
using the fact that
\begin{eqnarray}\label{partxT}
\partial_{\halpha}[Vaf']\hbA^{\halpha\bbeta}=\partial_y[V(T)af']\hbA^{y\bbeta}+
\partial_{\balpha}[V af']\hbA^{\balpha\bbeta}=
\partial_y[V(T)af']\partial_{\bdelta}t \ba^{\bdelta \bbeta}+
\partial_{\balpha}[V(T)af']\ba^{\balpha\bbeta}=0 \ . \nonumber \\
\end{eqnarray}
Further, we have  $f'>0$ by presumption. On the other hand since $V\sim e^{-T^2}$ we have
\begin{equation}
\lim_{a\rightarrow \infty}af'V\sim
\lim_{a\rightarrow \infty} a e^{-a^2(y-t)^2}=0
\end{equation}
for $y\neq t(\xi)$. So that the equation of motion (\ref{eqetapre}) is obeyed for
$t\neq y$ in the limit $a\rightarrow \infty$ while in case $t=y$ it is obeyed on condition when
\begin{equation}
\partial_{\balpha}\left[\frac{e^{-\phi}\sqrt{\det\ba}\ba^{\balpha\bbeta}_S\partial_{\bbeta}\eta}{\sqrt{(\tau_{\balpha}-\partial_{\balpha}\eta)\ba^{\balpha\bbeta}
		(\tau_{\bbeta}+\partial_{\bbeta}\eta)}}\right]=0
\end{equation}
which is precisely the equations of motion for non-relativistic D(p-1)-brane. To see this in more detail  let us consider an action for non-relativistic D(p-1)-brane that has the form
\cite{Kluson:2019avy}
\begin{equation}
S=-\ttau_{p-1}\int d^p\xi e^{-\phi}\sqrt{\det \ba}
\sqrt{(\tau_{\balpha}-\partial_{\balpha}\eta)\ba^{\balpha\bbeta}
(\tau_{\bbeta}+\partial_{\bbeta}\eta)} \ .
\end{equation}
From this action we get following equations of motion
\begin{eqnarray}\label{eqetabps}
\partial_{\balpha}\left[\frac{e^{-\phi}\sqrt{\det\ba}\ba^{\balpha
		\bbeta}_S\partial_{\bbeta}\eta}
{\sqrt{(\tau_{\balpha}-\partial_{\balpha}\eta)\ba^{\balpha\bbeta}(\tau_{\bbeta}+
		\partial_{\bbeta}\eta)}}\right]=0 \ , \nonumber \\
\end{eqnarray}
Further, equations of motion with respect to $a_{\balpha}$ take the form
\begin{eqnarray}\label{eqabps}
& &\partial_{\bbeta}[e^{-\phi}\ba^{\balpha\bbeta}_A\sqrt{\det\ba}
\sqrt{(\tau_{\balpha}-\partial_{\balpha}\eta)\ba^{\balpha\bbeta}
	(\tau_{\bbeta}+\partial_{\bbeta}\eta)}]-
\nonumber \\
& &-\frac{1}{2}\partial_{\bbeta}\left[e^{-\phi}\frac{\sqrt{\det\ba}}
{\sqrt{(\tau_{\balpha}-\partial_{\balpha}\eta)\ba^{\balpha\bbeta}
		(\tau_{\bbeta}+\partial_{\bbeta}\eta)}}	
(\tau_{\bgamma}-\partial_{\bgamma}\eta)\ba^{\bgamma\bbeta}\ba^{\balpha\bdelta}
(\tau_{\bdelta}+\partial_{\bdelta}\eta)\right]+\nonumber \\
& &+\frac{1}{2}\partial_{\bbeta}\left[e^{-\phi}\frac{\sqrt{\det\ba}}{
	\sqrt{(\tau_{\balpha}-\partial_{\balpha}\eta)\ba^{\balpha\bbeta}
		(\tau_{\bbeta}+\partial_{\bbeta}\eta)}}
(\tau_{\bgamma}-\partial_{\bgamma}\eta)\ba^{\bgamma\balpha}
\ba^{\bbeta\bdelta}(\tau_{\bdelta}+\partial_{\bdelta}\eta)\right]=0 \ .
\nonumber \\
\end{eqnarray}
Finally we determine equation of motion for $x^\mu$
\begin{eqnarray}\label{eqxbps}
& &\partial_\mu\phi e^{-\phi}
\sqrt{\det\ba}
\sqrt{(\tau_{\balpha}-\partial_{\balpha}\eta)\ba^{\balpha\bbeta}
	(\tau_{\bbeta}+\partial_{\bbeta}\eta)}-\nonumber \\
& &-\frac{1}{2} e^{-\phi}
\partial_\mu \bh_{\rho\sigma}\partial_{\balpha}x^\rho
\partial_{\bbeta}x^\sigma \ba^{\bbeta\balpha}\sqrt{\det\ba}
\sqrt{(\tau_{\balpha}-\partial_{\balpha}\eta)\ba^{\balpha\bbeta}
	(\tau_{\hbeta}+\partial_{\hbeta}\eta)}
+\nonumber \\
& &+\partial_{\balpha}\left[ e^{-\phi}
\bh_{\mu\nu}\partial_{\bbeta}x^\nu \ba^{\bbeta\balpha}\sqrt{\det\ba}
\sqrt{(\tau_{\balpha}-\partial_{\balpha}\eta)\ba^{\balpha\hbeta}
	(\tau_{\hbeta}+\partial_{\hbeta}\eta)}\right]+\nonumber \\
& &+\partial_{\balpha}\left[e^{-\phi}\sqrt{\det\ba}\tau_\mu
\frac{\ba^{\balpha\bbeta}_S\tau_{\bbeta}+\ba^{\balpha\bbeta}_A\partial_{\bbeta}\eta}{
	\sqrt{(\tau_{\balpha}-\partial_{\balpha}\eta)\ba^{\balpha\bbeta}
		(\tau_{\bbeta}+\partial_{\bbeta}\eta)}}\right] \nonumber \\
& &+\frac{1}{2} e^{-\phi}\frac{\sqrt{\det\ba}}
{\sqrt{(\tau_{\balpha}-\partial_{\balpha}\eta)\ba^{\balpha\bbeta}
		(\tau_{\bbeta}+\partial_{\bbeta}\eta)}}
(\tau_{\bgamma}-\partial_{\bgamma}\eta)\ba^{\bgamma\balpha}
\partial_\mu \bh_{\rho\sigma}\partial_{\balpha}x^\rho
\partial_{\bbeta}x^\sigma \ba^{\bbeta\bdelta}(\tau_{\bdelta}+
\partial_{\bdelta}\eta)- \nonumber \\
& &-\frac{1}{2}\partial_{\bgamma}\left[
e^{-\phi}\frac{\sqrt{\det\ba}}{\sqrt{(\tau_{\balpha}-\partial_{\balpha}\eta)\ba^{\balpha\bbeta}
		(\tau_{\bbeta}+\partial_{\bbeta}\eta)}}(\tau_{\balpha}-\partial_{\balpha}\eta)
\ba^{\balpha\bgamma}\bh_{\mu\nu}\partial_{\bdelta} x^\nu \ba^{\bdelta
	\hbeta}(\tau_{\hbeta}+\partial_{\hbeta}\eta)\right]-\nonumber \\
& &-\frac{1}{2}\partial_{\bdelta}\left[
e^{-\phi}\frac{\sqrt{\det\ba}}{\sqrt{(\tau_{\balpha}-\partial_{\balpha}\eta)\ba^{\balpha\bbeta}	(\tau_{\bbeta}+\partial_{\bbeta}\eta)}}(\tau_{\balpha}-\partial_{\balpha}\eta)
\ba^{\balpha\bgamma}\partial_{\bgamma}x^\nu\bh_{\nu\mu} \ba^{\bdelta
	\bbeta}(\tau_{\bbeta}+\partial_{\bbeta}\eta)\right]=0 \ . \nonumber  \\
\end{eqnarray}
Let us now return to the equations of motion for unstable D-brane. Inserting
(\ref{ans}) into  (\ref{eqT}) we obtain that is zero when
we take (\ref{partxT}) into account. This fact implies that $t(\xi)$
is not determined by equations of motion which  is a reflection of the fact
that $t(\xi)$ determines location of the kink on the world-volume of Dp-brane. On the other hand all positions of the kink on the world-volume of Dp-brane
are equivalent and hence  $t(\xi)$
should  be considered as parameter of diffeomorphism transformation rather then dynamical variable.

Let us further  consider equation of motion for $A_{\balpha}$ that for an ansatz
(\ref{ans}) takes the form
\begin{eqnarray}
& &af' V\left[\partial_{\bbeta}[e^{-\phi}\ba^{\balpha\bbeta}_A\sqrt{\det\ba}
\sqrt{(\tau_{\balpha}-\partial_{\balpha}\eta)\ba^{\balpha\bbeta}
	(\tau_{\bbeta}+\partial_{\bbeta}\eta)}]\right.-\nonumber \\
& &-\frac{1}{2}\partial_{\bbeta}[e^{-\phi}\frac{\sqrt{\det\ba}}
{\sqrt{(\tau_{\balpha}-\partial_{\balpha}\eta)\ba^{\balpha\bbeta}
		(\tau_{\bbeta}+\partial_{\bbeta}\eta)}}	
(\tau_{\bgamma}-\partial_{\bgamma}\eta)\ba^{\bgamma\bbeta}\ba^{\balpha\bdelta}
(\tau_{\bdelta}+\partial_{\bdelta}\eta)]+\nonumber \\
& & \left.+\frac{1}{2}\partial_{\bbeta}[e^{-\phi}\frac{\sqrt{\det\ba}}{
	\sqrt{(\tau_{\balpha}-\partial_{\balpha}\eta)\ba^{\balpha\bbeta}
		(\tau_{\bbeta}+\partial_{\bbeta}\eta)}}
(\tau_{\bgamma}-\partial_{\bgamma}\eta)\ba^{\bgamma\balpha}
\ba^{\bbeta\bdelta}(\tau_{\bdelta}+\partial_{\bdelta}\eta)\right]=0 \ ,  \nonumber \\
\end{eqnarray}
where we used (\ref{partxT}). We see that this equation is obeyed at the point $y=t(\xi)$   on condition that the expression in the bracket is zero. Comparing this
requirement with (\ref{eqabps}) we see that this is equation of motion for $a_{\balpha}$. In case of the equation of motion for $A_x$ we find
that it is again obeyed on condition that (\ref{eqabps}) holds. This follows from the fact that $\partial_{\balpha}[af' V]\hbA^{\balpha x}_A=
-\frac{d}{dx}[af'V]\partial_{\balpha}t\ba^{\balpha\bbeta}_A\partial_{\bbeta}t=0$.

 Finally we consider equation of motion for $x^\mu$. Following the same manipulation  and
 using (\ref{partxT}) we find that they hold at the point $y=t(\xi)$ on condition
 that the equations (\ref{eqxbps}) are obeyed.

In other words, we have shown that the tachyon kink solution on
non-relativistic non-BPS Dp-brane in torsional NC background can be identified as
stable non-relativistic D(p-1)-brane. This is similar situation as in case of relativistic non-BPS Dp-brane and tachyon kink solution.

\section{Hamiltonian for unstable Non-Relativistic  Dp-brane}\label{fourth}
In this section we derive Hamiltonian for unstable Dp-brane in torsional NC background
and study tachyon vacuum solution, generalization of the analysis presented in
\cite{Kluson:2019avy} to the case of unstable Dp-branes. We start with the well known Hamiltonian for relativistic unstable D(p+1)-brane that has the form
\begin{eqnarray}
& &\mH_{non}=N^\tau \mH_\tau^{non}+N^i\mH_i^{non}-A_0\mG \ , \nonumber \\
& &\mH_\tau^{non}=\Pi_M G^{MN}\Pi_N+l_s^{-4}\pi^i\partial_i x^M G_{MN}\partial_j x^N\pi^ j+l_s^{-2}
 \pi^i\partial_i T\partial_j T\pi^j+p_T^2+\nonumber \\
& & T_{p+1}^2 e^{-2\phi'}V(T)^2 \det( g_{ij}+l_s^2\partial_i T\partial_j T+l_s^2
F_{ij})\approx 0 \ ,
\nonumber \\
& &\mH_i^{non}=p_T\partial_i T+p_M\partial_i x^M+F_{ij}\pi^j\approx 0 \ , \quad \mG=\partial_i\pi^i\approx 0 \ ,  \nonumber \\
\end{eqnarray}
where $\Pi_M=p_M+l_s^{-2}B_{MN}\partial_i x^N\pi^i \ , i,j=1,\dots,p+1$ and where
$p_M$ are momenta conjugate to $x^M$, $\pi^i$ are momenta conjugate to $A_i$ and
$\mH_\tau^{non},\mH_i^{non}$ and $\mG$ are first class constraints.

Now we consider this unstable D(p+1)-brane in the null background given in
(\ref{Nullback}) and presume that it is extended along $u-$direction
so that
\begin{equation}\label{gaugefix}
u=\xi^{p+1}
\end{equation}
and all fields do not depend on $\xi^p$. Let $\hi,\hj=1,\dots,p$ denote
remaining spatial coordinates. Further we presume that components of NSNS two form are zero.  We can also interpret (\ref{gaugefix}) as the gauge fixing constraint
that together with $\mH_u$ are two second class constraints that can be explicitly solved so that
\begin{equation}
p_u=\partial_{\hi}A_u\pi^{\hi} \ .
\end{equation}
Then, following \cite{Kluson:2019avy},  we identify $A_u$ with $\eta$ and conjugate momentum $\pi^u$ with $p_\eta$ as
\begin{equation}
\eta=l_s^2 A_u \ , \quad p_\eta=l_s^{-2}\pi^u \ .
\end{equation}
As a result we obtain   Hamiltonian and spatial diffeomorphism constraints for unstable non-relativistic Dp-brane in the form
\begin{eqnarray}\label{Hnon}
& &\mH^{non}_\tau=2l_s^{-4}\partial_{\hi}\eta \pi^{\hi}\Phi \partial_{\hj}\eta \pi^{\hj}-2l_s^{-2}
\partial_{\hi}\eta \pi^{\hi}\hv^\mu p_\mu+p_\mu h^{\mu\nu}p_\nu+\nonumber \\
& &+l_s^{-4}\pi^{\hi}\partial_{\hi}x^\mu \bh_{\mu\nu}\partial_{\hj}x^\nu \pi^{\hj}+
l_s^{-2}\pi^{\hi}\partial_{\hi}T \partial_{\hj}T\pi^{\hj}+p_T^2
+2l_s^{-2}\pi^{\hi}\partial_{\hi}x^\mu \tau_\mu p_\eta+\nonumber \\
& &+T_p^2 e^{-2\phi}V(T)(\partial_{\hi}\eta \ba^{\hi \hj}
\partial_{\hj}\eta-\partial_{\hi} x^\mu \tau_\mu
\ba^{\hi\hj}\partial_{\hj}x^\nu\tau_\nu)\det \ba_{\hi \hj} \ ,
\nonumber \\
& &\mH^{non}_{\hi}=p_T\partial_{\hi}T+p_\mu\partial_{\hi}x^\mu+p_\eta\partial_{\hi}\eta+F_{\hi\hj}\pi^{\hj} \ , \quad \mG^{non}=\partial_{\hi}\pi^{\hi} \ , \nonumber \\
\end{eqnarray}
where
\begin{equation}
\ba_{\hi\hj}=\bh_{\hi\hj}+l_s^2\partial_{\hi}T\partial_{\hj}T+l_s^2
F_{\hi \hj} \  ,
\end{equation}
and where we performed rescaling as in \cite{Kluson:2019avy}. Alternatively, we could also derive (\ref{Hnon}) from (\ref{actnon}) through Legendre transformation but it is clear that these two procedures are equivalent.

As the next step we determine  canonical  equations of motion. Since $H^{non}=\int d^p\xi
(N^\tau\mH^{non}_\tau+N^{\hi}\mH^{non}_{\hi}-A_0\mG^{non})$ we obtain
\begin{eqnarray}\label{eqetacan}
& &\partial_\tau \eta=\pb{\eta,H^{non}}=2N^\tau l_s^{-2}\pi^{\hi}\partial_{\hi}x^\mu\tau_\mu+
N^{\hi}\partial_{\hi}\eta  \ , \nonumber \\
& &\partial_\tau p_\eta=\pb{p_\eta,H^{non}}=
\partial_{\hi}[4l_s^{-4}N^\tau \pi^{\hi}\Phi\partial_{\hj}\eta \pi^{\hj}-2l_s^{-2}N^\tau
\pi^{\hi}\hv^\mu p_\mu +2T_p^2N^\tau e^{-2\phi}V \ba^{\hi\hj}\partial_{\hj}\eta \det\ba+
N^\sigma p_\eta] \  \nonumber \\
\end{eqnarray}
while equations of motion for $T$ and $p_T$ have the form
\begin{eqnarray}\label{eqTcan}
& &\partial_\tau T=\pb{T,H^{non}}=2N^\tau p_T+N^{\hi}\partial_{\hi}T \ , \nonumber \\
& &\partial_\tau p_T=\pb{p_T,H^{non}}=\partial_{\hi}[2N^\tau l_s^{-2}\pi^{\hi}\partial_{\hj}T
\pi^{\hj}]+\partial_{\hi}[N^{\hi}p_T]-
\nonumber \\
& &-T_p^2 e^{-2\phi}\frac{dV(T)}{dT}(\partial_{\hi}\eta \ba^{\hi \hj}
\partial_{\hj}\eta-\partial_{\hi} x^\mu \tau_\mu
\ba^{\hi\hj}\partial_{\hj}x^\nu\tau_\nu)\det \ba-
\nonumber \\
& &-\partial_{\hi}[T_p^2 e^{-2\phi}V(T)\det\ba(\partial_{\hk}\eta\ba^{\hk\hi}\partial_{\hj}T\ba^{\hj\hl}\partial_{\hl}\eta+
\partial_{\hk}\eta \ba^{\hk\hj}\partial_{\hj}T\ba^{\hi\hl}\partial_{\hl}\eta)]+\nonumber \\
& &+\partial_{\hi}[T_p^2 e^{-2\phi}V(T)\det\ba(\partial_{\hk}x^\mu\tau_\mu\ba^{\hk\hi}\partial_{\hj}T\ba^{\hj\hl}\partial_{\hl}x^\nu\tau_\nu+
\partial_{\hk}x^\mu\tau_\mu \ba^{\hk\hj}\partial_{\hj}T\ba^{\hi\hl}\partial_{\hl}x^\nu\tau_\nu)]+
\nonumber \\
& &+2\partial_{\hi}[N^\tau T^2_p e^{-2\phi} V\det\ba
\partial_{\hj}T \ba^{\hj \hi}_S(\partial_{\hi}\eta \ba^{\hi \hj}
\partial_{\hj}\eta-\partial_{\hi} x^\mu \tau_\mu
\ba^{\hi\hj}\partial_{\hj}x^\nu\tau_\nu)] \ .  \nonumber \\
\end{eqnarray}
Further, the equations of motion for $x^\mu$ and $p_\mu$  have the form
\begin{eqnarray}\label{eqxcan}
& &\partial_\tau x^\mu=\pb{x^\mu,H^{non}}=-2N^\tau l_s^{-2}\partial_{\hi}\eta\pi^{\hi}\hv^\mu+
2N^\tau h^{\mu\nu}p_\nu+N^{\hi}\partial_{\hi}x^\mu \  ,
\nonumber \\
& &\partial_\tau p_\mu=\pb{p_\mu,H}=-2N^\tau l_s^{-4}\partial_{\hi}\eta \pi^{\hi}
\partial_\mu\Phi\partial_{\hj}\eta \pi^{\hj}+2N^\tau l_s^{-2}\partial_{\hi}\eta \pi^{\hi}
\partial_\mu \hv^\nu p_{\nu}-2N^\tau p_\rho \partial_\mu h^{\rho\sigma}p_\nu+\nonumber \\
& &+2N^\tau l_s^{-4}\partial_{\hi}[\pi^{\hi} \bh_{\mu\nu}\partial_{\hj}x^\nu\pi^{\hj}]-
2N^\tau l_s^{-4}\pi^{\hi}\partial_{\hi}x^\rho\partial_\mu \bh_{\rho\sigma}\partial_{\hj}x^\nu
\pi^{\hj}-\nonumber \\
& &-\frac{2}{l_s^2}\partial_{\hi}[N^\tau \pi^{\hi}\tau_\mu p_\eta ]-2N^\tau l_s^{-2}\pi^{\hi}\partial_{\hi}x^\nu\partial_\mu \tau_\nu p_\eta+\nonumber \\
& &+T_p^2  N^\tau\partial_\mu [e^{-2\phi}]V
(\partial_{\hi}\eta \ba^{\hi \hj}
\partial_{\hj}\eta-\partial_{\hi} x^\mu \tau_\mu
\ba^{\hi\hj}\partial_{\hj}x^\nu\tau_\nu)\det \ba+
\nonumber \\
& &+T_p^2 N^\tau e^{-2\phi}V[\partial_{\hk}\eta \ba^{\hk \hi}\partial_\mu \bh_{\rho\sigma}
\partial_{\hi}x^\rho\partial_{\hj}x^\sigma \ba^{\hj\hl}\partial_{\hl}\eta
-\partial_{\hk}x^{\rho'}\tau_{\rho'}\ba^{\hk \hi}\partial_\mu \bh_{\rho\sigma}
\partial_{\hi}x^\rho\partial_{\hj}x^\sigma \ba^{\hj\hl}\partial_{\hl}x^{\sigma'}\tau_{\sigma'}]\det\ba-\nonumber \\
& &-\partial_{\hi}[T_p^2 N^\tau e^{-2\phi}V\det\ba
\partial_{\hk}\eta \ba^{\hk\hi}\bh_{\mu\nu}\partial_{\hj}x^\nu
\ba^{\hj\hl}\partial_{\hl}\eta]
-\partial_{\hj}[T_p^2N^\tau e^{-2\phi}V\det \ba
\partial_{\hk}\eta \ba^{\hk\hi}\bh_{\nu\mu}\partial_{\hi}x^\nu
\ba^{\hj\hl}\partial_{\hl}\eta]+\nonumber \\
& &+\partial_{\hi}[T_p^2 N^\tau e^{-2\phi}V\det\ba
\tau_{\hk} \ba^{\hk\hi}\bh_{\mu\nu}\partial_{\hj}x^\nu
\ba^{\hj\hl}\tau_{\hl}]
-\partial_{\hj}[T_p^2N^\tau e^{-2\phi}V\det \ba
\tau_{\hk} \ba^{\hk\hi}\bh_{\nu\mu}\partial_{\hi}x^\nu
\ba^{\hj\hl}\tau_{\hl}]-\nonumber \\
& &-2T_p^2\partial_{\hi}[N^\tau e^{-2\phi}V\tau_\mu\ba^{\hi\hj}
\tau_{\hj}\det\ba]+\nonumber \\
& &-T_p^2 N^\tau e^{-2\phi}V(\partial_{\hi}\eta\ba^{\hi\hj}
\partial_{\hj}\eta-\tau_{\hi}\ba^{\hi\hj}\tau_{\hj})
\partial_\mu \bh_{\rho\sigma}\partial_{\hk}x^\rho\partial_{\hl}x^\sigma
\ba^{\hl\hk}\det\ba+\nonumber \\
& &+2\partial_{\hi}[N^\tau T_p^2 e^{-2\phi}
V(\partial_{\hk}\eta\ba^{\hk\hl}\partial_{\hl}\eta-
\tau_{\hk}\ba^{\hk\hl}\tau_{\hl})\bh_{\mu\nu}\partial_{\hj}x^\nu
\ba^{\hj\hi}_S\det\ba]+\partial_{\hi}[N^{\hi}p_\mu] \ . \nonumber \\
\end{eqnarray}
Finally we determine equations of motion for $A_{\hi}$ and $\pi^{\hi}$
\begin{eqnarray}
& &\partial_\tau A_{\hi}=\pb{A_{\hi},H^{non}}=4N^\tau l_s^{-4}\partial_{\hi}\eta
\Phi\partial_{\hj}\eta \pi^{\hj}-2N^\tau l_s^{-2}\partial_{\hi}\eta \hv^\mu p_\mu+
\nonumber \\
& &+2N^\tau l_s^{-4}\partial_{\hi}x^\mu \bh_{\mu\nu}\partial_{\hj}x^\nu\pi^{\hj}+2N^\tau
l_s^{-2}\partial_{\hi}T\partial_{\hj}T\pi^{\hj}+2N^\tau l_s^{-2}\partial_{\hi}x^\mu
\tau_\mu p_\eta+N^{\hj}F_{\hj\hi}+\partial_{\hi}A_0 \ , \nonumber \\
& &\partial_\tau \pi^{\hi}=\pb{\pi^{\hi},H^{non}}=\partial_{\hk}(N^{\hk}\pi^{\hi})-\partial_{\hl}(N^{\hi}\pi^{\hl})-
T_p^2\partial_{\hj}[ N^\tau e^{-2\phi}V\partial_{\hk}\eta \ba^{\hk\hj}\ba^{\hi\hl}\partial_{\hl}\eta \det\ba
\nonumber \\
& &+\partial_{\hj}[N^\tau e^{-2\phi}V N^\tau e^{-2\phi}V\partial_{\hk}\eta \ba^{\hk\hi}\ba^{\hj\hl}\partial_{\hl}\eta \det\ba]
\nonumber \\
&& +T_p^2\partial_{\hj}[ N^\tau e^{-2\phi}V\tau_{\hk} \ba^{\hk\hj}\ba^{\hi\hl}\tau_{\hl} \det\ba
\nonumber \\
& &-\partial_{\hj}[N^\tau e^{-2\phi}V N^\tau e^{-2\phi}V\tau_{\hk} \ba^{\hk\hi}\ba^{\hj\hl}\tau_{\hl}\eta \det\ba]
\nonumber \\
& &+2l_s^2T_p^2 \partial_{\hj}[N^\tau e^{-2\phi}V
(\partial_{\hk}\eta \ba^{\hk\hl}\partial_{\hl}\eta-\tau_{\hk}\ba^{\hk\hl}
\tau_{\hl})\ba^{\hi\hj}_A\det\ba] \ .
\nonumber \\
\end{eqnarray}
We see that these equations of motion are very complicated. On the other hand we will be interested in the tachyon vacuum solution $T=T_{min},p_T=0$ that correspond to $V(T_{min})=0 \ ,
\frac{dV}{dT}(T_{min})=0$. Further, $\partial_{\hi} V(T_{min})=\frac{dV}{dT}(T_{min})\partial_{\hi} T=0$ since we have tachyon vacuum solution everywhere on the world-volume on non-relativistic unstable Dp-brane.  In this case the equation of motion simplify considerably
\begin{eqnarray}
& & \partial_\tau \eta=2N^\tau l_s^{-2}\pi^{\hi}\partial_{\hi}x^\mu\tau_\mu+
 N^{\hi}\partial_{\hi}\eta  \ , \nonumber \\
& & \partial_\tau p_\eta=
 \partial_{\hi}[4l_s^{-4}N^\tau \pi^{\hi}\Phi\partial_{\hj}\eta \pi^{\hj}-2l_s^{-2}N^\tau
 \pi^{\hi}\hv^\mu p_\mu +
 N^\sigma p_\eta] \  \nonumber \\
 \end{eqnarray}
and
\begin{eqnarray}
& &\partial_\tau x^\mu=-2N^\tau l_s^{-2}\partial_{\hi}\eta\pi^{\hi}\hv^\mu+
2N^\tau h^{\mu\nu}p_\nu+N^{\hi}\partial_{\hi}x^\mu \  ,
\nonumber \\
& &\partial_\tau p_\mu=-2l_s^{-4}\partial_{\hi}\eta \pi^{\hi}
\partial_\mu\Phi\partial_{\hj}\eta \pi^{\hj}+2N^\tau l_s^{-2}\partial_{\hi}\eta \pi^{\hi}
\partial_\mu \hv^\nu p_{\nu}-2N^\tau p_\rho \partial_\mu h^{\rho\sigma}p_\nu+\nonumber \\
& &+2N^\tau l_s^{-4}\partial_{\hi}[\pi^{\hi} \bh_{\mu\nu}\partial_{\hj}x^\nu\pi^{\hj}]-
2N^\tau l_s^{-4}\pi^{\hi}\partial_{\hi}x^\rho\partial_\mu \bh_{\rho\sigma}\partial_{\hj}x^\nu
\pi^{\hj}-\nonumber \\
& &-\frac{2}{l_s^2}\partial_{\hi}[N^\tau \pi^{\hi}\tau_\mu p_\eta ]
+2N^\tau l_s^{-2}\pi^{\hi}\partial_{\hi}x^\nu\partial_\mu \tau_\nu p_\eta+\partial_{\hi}[N^{\hi}p_\mu] \ . \nonumber \\
\end{eqnarray}
Finally the equations of motion for $A_{\hi}$ and $\pi^{\hi}$ have the form
\begin{eqnarray}\label{eqAvac}
& &\partial_\tau A_{\hi}=4N^\tau l_s^{-4}\partial_{\hi}\eta
\Phi\partial_{\hj}\eta \pi^{\hj}-2N^\tau l_s^{-2}\partial_{\hi}\eta \hv^\mu p_\mu+\partial_{\hi}A_0
\nonumber \\
& &+2N^\tau l_s^{-4}\partial_{\hi}x^\mu \bh_{\mu\nu}\partial_{\hj}x^\nu\pi^{\hj}+2N^\tau l_s^{-2}\partial_{\hi}x^\mu
\tau_\mu p_\eta+N^{\hj}F_{\hj\hi} \ , \nonumber \\
& &\partial_\tau \pi^{\hi}=\partial_{\hk}(N^{\hk}\pi^{\hi})-
\partial_{\hk}(N^{\hi}\pi^{\hk}) \ .
\nonumber \\
\end{eqnarray}
To proceed further we introduce projector on direction along $\pi^{\hi}$
\begin{equation}
\triangle^{\hi}_{ \ \hj}=\delta^{\hi}_{\hj}-
\frac{\pi^{\hi}\pi^{\hl}\bh_{\hl\hj}}{
\pi^{\hk}\bh_{\hk\hl}\pi^{\hl}} \ , \quad  \bh_{\hi\hj}=
\partial_{\hi}x^\mu \bh_{\mu\nu}\partial_{\hj}x^\nu \ , \quad
\triangle^{\hi}_{ \ \hj}\triangle^{\hj}_{ \ \hk}=
\triangle^{\hi}_{ \ \hk} \ .
\end{equation}
With he help of this projector we can split $N^{\hi}$ into two components as
\begin{equation}
N^{\hi}=N^{\hk}\delta_{\hk}^{\hi}=
\triangle^{\hi}_{ \ \hk}N^{\hk}+\pi^{\hi}\frac{\pi^{\hl}\bh_{\hl\hk}
	N^{\hk}}{\pi^{\hl}\bh_{\hl\hk}\pi^{\hk}}\equiv
N^{\hi}_{T}+N^{\hi}_{II}\pi^{\hi} \ ,
\end{equation}
where
\begin{equation}
N^{\hi}_T\bh_{ij}\pi^{\hj}=0 \ .
\end{equation}
Let us consider solutions where $\pi^{\hi}$ is constant and hence
$\partial_\tau \pi^{\hi}=0 \ , \quad  \partial_{\hj}\pi^{\hi}=0$ which also implies that
the Gauss constraints is obeyed.

Then the second equation in (\ref{eqAvac}) takes the form
\begin{equation}
0=\partial_{\hk}N^{\hk}_T\pi^{\hi}-\partial_{\hk}N^{\hi}_T\pi^{\hk}
\end{equation}
that should be valid for all $\hi$. This can be obeyed on condition when
\begin{equation}
N^{\hi}_T=0 \ .
\end{equation}
 Further, we know that the physical dimension of $\pi^{\hi}$ is $[\pi^{\hi}]=L^{-(p+1)}$ where $L$ is some length scale. Then we can write
$\pi^{\hi}$ as
\begin{equation}
\pi^{\hi}=n^{\hi}T_p \ ,
\end{equation}
where $n^{\hi}$ is dimensionless. Since $\pi^{\hi}$ always appears with combination
$\pi^{\hi}\partial_{\hi}$ we introduce coordinate $\sigma$ in the following way:
\begin{equation}
\pi^{\hi}\partial_{\hi}=T_pn^{\hi}\partial_{\hi}\equiv T_p \partial_\sigma \ .
\end{equation}
Let us perform some dimensional analysis and rescaling. First of all we have $N^{\hi}=N_{II}T_p n^{\hi}$. Further  $N^\tau$ has dimension $[N^\tau]=L^{p+1}$ and hence it can be written as $N^\tau=\frac{n^\tau}{T_p}$ where $n^\tau$ is dimensionless. Further, since
$N^{\hi}$ is dimensionless  we find that $N_{II}$ can be written as
\begin{equation}
N_{II}=n^\sigma \frac{1}{T_p} \ ,
\end{equation}
where $n^\sigma$ is dimensionless. Finally we write all momenta as
\begin{equation}
p_\eta=k_\eta \frac{T_p}{l_s^2} \ , \quad  p_\mu=k_\mu\frac{T_p}{l_s^2} \ .
\end{equation}
Using these rescaled variables we get the equation of motion in the form
\begin{eqnarray}\label{eqetafin}
& &\partial_\tau \eta=2n^\tau l_s^{-2}\partial_\sigma x^\mu \tau_\mu+n^\sigma
\partial_\sigma \eta \ , \nonumber \\
& &\partial_\tau k_\eta=\frac{1}{l_s^2}
\partial_\sigma[4n^\tau \Phi\partial_\sigma \eta-2n^\tau \hv^\mu k_\mu
-n^\sigma k_\eta] \ , \nonumber \\
\end{eqnarray}
while equations of motion for $x^\mu$ and $p_\mu$ reduce into
\begin{eqnarray}\label{eqxmufin}
& &\partial_\tau x^\mu=-\frac{2}{l_s^2}n^\tau\partial_\sigma \eta\hv^\mu+2
n^\tau h^{\mu\nu}p_\nu+n^\sigma \partial_\sigma x^\mu \ , \nonumber \\
& &\partial_\tau k_\mu=-\frac{2n^\tau}{l_s^2}\partial_\sigma\eta \partial_\mu\Phi
\partial_\sigma \eta+\frac{2n^\tau}{l_s^2}\partial_\sigma \eta \partial_\mu
\hv^\nu k_\nu-\nonumber \\
& &-2\frac{n^\tau}{l_s^2}k_\rho \partial_\mu h^{\rho\sigma}k_\sigma+\frac{2n^\tau}{l_s^2}\partial_\sigma[\bh_{\mu\nu}
\partial_\sigma x^\nu]-\frac{2n^\tau}{l_s^2}\partial_\sigma x^\rho
\partial_\mu \bh_{\rho\sigma}\partial_\sigma x^\nu-\nonumber \\
& &-2\frac{n^\tau}{l_s^2}\partial_\sigma x^\nu\partial_\mu \tau_\nu k_\eta+
n^\sigma\partial_\sigma x^\mu+\frac{2}{l_s^2}\partial_
\sigma[n^\tau \tau_\mu k_\eta] \ .\nonumber \\
\end{eqnarray}
Now we compare these equations with the equations of motion for non-relativistic string
in torsional NC background whose Hamiltonian was derived in \cite{Kluson:2018egd} in the form
\begin{equation}\label{stringHam}
K=\int d\sigma (n^\tau \mK_\tau+n^\sigma \mK_\sigma) \  ,
\end{equation}
where
\begin{eqnarray}
& &\mK_\tau=k_\mu h^{\mu\nu}k_\nu-\frac{2}{l_s^2}k_\mu \hv^\mu\partial_\sigma \eta+
\frac{1}{l_s^4}\bh_{\mu\nu}\partial_\sigma x^\mu\partial_\sigma x^\nu+
\frac{2}{l_s^2}k_\eta \tau_\mu\partial_\sigma x^\mu+
\frac{2}{l_s^4}\partial_\sigma\eta \Phi\partial_\sigma \eta \ ,
\nonumber \\
& &\mH_\sigma=k_\mu\partial_\sigma x^\mu \ .
\nonumber \\
\end{eqnarray}
Using string Hamiltonian (\ref{stringHam}) we obtain following
 equations of motion for non-relativistic string in torsional NC background in the form
\begin{eqnarray}\label{eqstringfun}
& &	\partial_\tau x^\mu=\pb{x^\mu,K}=n^\tau (2h^{\mu\nu}k_\nu-\frac{2}{l_s^2}
	\hv^\mu\partial_\sigma \eta)+n^\sigma \partial_\sigma x^\mu \ , \nonumber \\
& &\partial_\tau k_\mu=\pb{p_\mu,K}=-n^\tau k_\rho \partial_\mu h^{\rho\sigma}k_\sigma
+\frac{2n^\tau}{l_s^2}k_\nu \partial_\mu \hv^\nu\partial_\sigma\eta-\frac{1}{l_s^2}
\partial_\mu \bh_{\rho\sigma}\partial_\sigma x^\rho\partial_\sigma x^\sigma+
\nonumber \\
& &+\frac{2}{l_s^4}\partial_\sigma[n^\tau \bh_{\mu\nu}\partial_\sigma x^\nu]-
\frac{2n^\tau}{l_s^2}k_\eta \partial_\mu \tau_\nu\partial_\sigma x^\nu+
\frac{2}{l_s^2}\partial_\sigma[k_\eta \tau_\mu]-\frac{2n^\tau}{l_s^4}
\partial_\sigma \eta \partial_\mu\Phi \partial_\sigma \eta+\partial_\sigma[n^\sigma k_\mu] \  \nonumber \\	
	\end{eqnarray}
	and also
\begin{eqnarray}\label{eqstringfun1}
& &\partial_\tau \eta=\pb{\eta,K}=\frac{2n^\tau}{l_s^2}\tau_\mu
\partial_\sigma x^\mu+n^\sigma\partial_\sigma \eta \ , \nonumber \\
& &\partial_\tau k_\eta=\pb{k_\eta,K}=
-\partial_\sigma[\frac{2n^\tau}{l_s^2}k_\mu\hv^\mu]+\frac{4}{l_s^4}
\partial_\sigma[n^\tau \Phi\partial_\sigma \eta]+\partial_\sigma[n^\sigma k_\eta] \ .
\nonumber \\
\end{eqnarray}
We see that these equations coincide with  (\ref{eqetafin}) and
(\ref{eqxmufin}) and hence we can interpret the tachyon vacuum solution as a state
with the gas of fundamental non-relativistic strings. More precisely, the equations
of motion (\ref{eqetafin}) and (\ref{eqxmufin}) contain derivative with respect to $\sigma$ while generally fields defined on the world-volume of unstable Dp-brane depend on
transverse coordinates (transverse with respect to $\sigma$) as well. In other words all solutions of (\ref{eqetafin}) and (\ref{eqxmufin}) can be multiplied by function that
depend on transverse coordinates and that can be interpreted as density of the fundamental non-relativistic  strings as in \cite{Sen:2003bc,Sen:2000kd}. We mean that this is again
nice consistency check of the proposed non-relativistic unstable Dp-brane so that it can
be considered as important part of non-relativistic string theory in torsional NC background.


\begin{thebibliography}{20}
	

\bibitem{Cartan:1923zea}
E.~Cartan,
\emph{``Sur les variétés à connexion affine et la théorie de la relativité généralisée. (première partie),''}
Annales Sci.\ Ecole Norm.\ Sup.\  {\bf 40} (1923) 325.

	
\bibitem{Christensen:2013lma}
M.~H.~Christensen, J.~Hartong, N.~A.~Obers and B.~Rollier,
\emph{``Torsional Newton-Cartan Geometry
	and Lifshitz Holography,''}
Phys.\ Rev.\ D {\bf 89} (2014) 061901
doi:10.1103/PhysRevD.89.061901
[arXiv:1311.4794 [hep-th]].

\bibitem{Christensen:2013rfa}
M.~H.~Christensen, J.~Hartong, N.~A.~Obers and B.~Rollier,
\emph{``Boundary Stress-Energy Tensor and Newton-Cartan Geometry in Lifshitz Holography,''}
JHEP {\bf 1401} (2014) 057
doi:10.1007/JHEP01(2014)057
[arXiv:1311.6471 [hep-th]].
	
	\bibitem{Hartong:2014oma}
	J.~Hartong, E.~Kiritsis and N.~A.~Obers,
\emph{"Lifshitz space–times for Schrödinger holography,''}
	Phys.\ Lett.\ B {\bf 746} (2015) 318
	doi:10.1016/j.physletb.2015.05.010
	[arXiv:1409.1519 [hep-th]].




\bibitem{Horava:2009uw}
P.~Horava,
\emph{``Quantum Gravity at a Lifshitz Point,''}
Phys.\ Rev.\ D {\bf 79} (2009) 084008
doi:10.1103/PhysRevD.79.084008
[arXiv:0901.3775 [hep-th]].



\bibitem{Hartong:2015zia}
J.~Hartong and N.~A.~Obers,
\emph{``Hořava-Lifshitz gravity
	from dynamical Newton-Cartan geometry,''}
JHEP {\bf 1507} (2015) 155
doi:10.1007/JHEP07(2015)155
[arXiv:1504.07461 [hep-th]].



\bibitem{Hartong:2016yrf}
J.~Hartong, Y.~Lei and N.~A.~Obers,
\emph{``Nonrelativistic Chern-Simons theories and three-dimensional Hořava-Lifshitz gravity,''}
Phys.\ Rev.\ D {\bf 94} (2016) no.6,  065027
doi:10.1103/PhysRevD.94.065027
[arXiv:1604.08054 [hep-th]].



\bibitem{Kluson:2018uss}
J.~Kluso\v{n},
\emph{"Hamiltonian for a string in a Newton-Cartan background,''}
Phys.\ Rev.\ D {\bf 98} (2018) no.8,  086010
doi:10.1103/PhysRevD.98.086010
[arXiv:1801.10376 [hep-th]].



%















\bibitem{Gomis:2000bd}
J.~Gomis and H.~Ooguri,
\emph{``Nonrelativistic closed string theory,''}
J.\ Math.\ Phys.\  {\bf 42} (2001) 3127
doi:10.1063/1.1372697
[hep-th/0009181].


\bibitem{Danielsson:2000gi}
U.~H.~Danielsson, A.~Guijosa and M.~Kruczenski,
\emph{``IIA/B, wound and wrapped,''}
JHEP {\bf 0010} (2000) 020
doi:10.1088/1126-6708/2000/10/020
[hep-th/0009182].




\bibitem{Andringa:2012uz}
R.~Andringa, E.~Bergshoeff, J.~Gomis and M.~de Roo,
\emph{``'Stringy' Newton-Cartan Gravity,''}
Class.\ Quant.\ Grav.\  {\bf 29} (2012) 235020
doi:10.1088/0264-9381/29/23/235020
[arXiv:1206.5176 [hep-th]].

\bibitem{Harmark:2017rpg}
T.~Harmark, J.~Hartong and N.~A.~Obers,
\emph{``Nonrelativistic
	strings and limits of the AdS/CFT correspondence,''}
Phys.\ Rev.\ D {\bf 96} (2017) no.8,  086019
doi:10.1103/PhysRevD.96.086019
[arXiv:1705.03535 [hep-th]].



\bibitem{Bergshoeff:2018yvt}
E.~Bergshoeff, J.~Gomis and Z.~Yan,
\emph{``Nonrelativistic String Theory and T-Duality,''}
JHEP {\bf 1811} (2018) 133
doi:10.1007/JHEP11(2018)133
[arXiv:1806.06071 [hep-th]].

\bibitem{Kluson:2018egd}
J.~Kluso\v{n},
\emph{``Remark About Non-Relativistic String in Newton-Cartan Background and Null Reduction,''}
JHEP {\bf 1805} (2018) 041
doi:10.1007/JHEP05(2018)041
[arXiv:1803.07336 [hep-th]].


\bibitem{Kluson:2018grx}
J.~Klusoň,
\emph{``Nonrelativistic String Theory
	Sigma Model and Its Canonical Formulation,''}
Eur.\ Phys.\ J.\ C {\bf 79} (2019) no.2,  108
doi:10.1140/epjc/s10052-019-6623-9
[arXiv:1809.10411 [hep-th]].

\bibitem{Harmark:2018cdl}
T.~Harmark, J.~Hartong, L.~Menculini, N.~A.~Obers and Z.~Yan,
\emph{``Strings with Non-Relativistic Conformal Symmetry and Limits of the AdS/CFT Correspondence,''}
JHEP {\bf 1811} (2018) 190
doi:10.1007/JHEP11(2018)190
[arXiv:1810.05560 [hep-th]].



\bibitem{Gomis:2019zyu}
J.~Gomis, J.~Oh and Z.~Yan,
\emph{``Nonrelativistic String Theory in Background Fields,''}
JHEP {\bf 1910} (2019) 101
doi:10.1007/JHEP10(2019)101
[arXiv:1905.07315 [hep-th]].

\bibitem{Gallegos:2019icg}
A.~D.~Gallegos, U.~Gürsoy and N.~Zinnato,
\emph{``Torsional Newton Cartan gravity from non-relativistic strings,''}
arXiv:1906.01607 [hep-th].

\bibitem{Bergshoeff:2019pij}
E.~A.~Bergshoeff, J.~Gomis, J.~Rosseel, C.~Şimşek and Z.~Yan,
\emph{``String Theory and String Newton-Cartan Geometry,''}
J.\ Phys.\ A {\bf 53} (2020) no.1,  014001
doi:10.1088/1751-8121/ab56e9
[arXiv:1907.10668 [hep-th]].






\bibitem{Harmark:2019upf}
T.~Harmark, J.~Hartong, L.~Menculini, N.~A.~Obers and G.~Oling,
\emph{``Relating non-relativistic string theories,''}
JHEP {\bf 1911} (2019) 071
doi:10.1007/JHEP11(2019)071
[arXiv:1907.01663 [hep-th]].

\bibitem{Bergshoeff:2015uaa}
E.~Bergshoeff, J.~Rosseel and T.~Zojer,
\emph{``Newton–Cartan (super)gravity as a non-relativistic limit,''}
Class.\ Quant.\ Grav.\  {\bf 32} (2015) no.20,  205003
doi:10.1088/0264-9381/32/20/205003
[arXiv:1505.02095 [hep-th]].


\bibitem{Kluson:2019avy}
J.~Klusoň,
\emph{``Non-Relativistic D-brane from T-duality Along Null Direction,''}
JHEP {\bf 1910} (2019) 153
doi:10.1007/JHEP10(2019)153
[arXiv:1907.05662 [hep-th]].

\bibitem{Simon:2011rw}
J.~Simon,
\emph{``Brane Effective Actions, Kappa-Symmetry and Applications,''}
Living Rev.\ Rel.\  {\bf 15} (2012) 3
doi:10.12942/lrr-2012-3
[arXiv:1110.2422 [hep-th]].

\bibitem{Sen:2004nf}
A.~Sen,
\emph{``Tachyon dynamics in open string theory,''}
Int.\ J.\ Mod.\ Phys.\ A {\bf 20} (2005) 5513
doi:10.1142/S0217751X0502519X
[hep-th/0410103].




\bibitem{Sen:1999md}
A.~Sen,
\emph{``Supersymmetric world volume action for nonBPS D-branes,''}
JHEP {\bf 9910} (1999) 008
doi:10.1088/1126-6708/1999/10/008
[hep-th/9909062].


\bibitem{Kluson:2000iy}
J.~Kluso\v{n},
\emph{``Proposal for nonBPS D-brane action,''}
Phys.\ Rev.\ D {\bf 62} (2000) 126003
doi:10.1103/PhysRevD.62.126003
[hep-th/0004106].


\bibitem{Bergshoeff:2000dq}
E.~A.~Bergshoeff, M.~de Roo, T.~C.~de Wit, E.~Eyras and S.~Panda,
\emph{``T duality and actions for nonBPS D-branes,''}
JHEP {\bf 0005} (2000) 009
doi:10.1088/1126-6708/2000/05/009
[hep-th/0003221].










\bibitem{Sen:2003tm}
A.~Sen,
\emph{``Dirac-Born-Infeld action on the tachyon kink and vortex,''}
Phys.\ Rev.\ D {\bf 68} (2003) 066008
doi:10.1103/PhysRevD.68.066008
[hep-th/0303057].



\bibitem{Sen:2003bc}
A.~Sen,
\emph{``Open and closed strings from unstable D-branes,''}
Phys.\ Rev.\ D {\bf 68} (2003) 106003
doi:10.1103/PhysRevD.68.106003
[hep-th/0305011].

\bibitem{Sen:2000kd}
A.~Sen,
\emph{``Fundamental strings in
	open string theory at the tachyonic vacuum,''}
J.\ Math.\ Phys.\  {\bf 42} (2001) 2844
doi:10.1063/1.1377037
[hep-th/0010240].

\end{thebibliography}
\end{document}